\documentclass[12pt]{article}
\usepackage{epsfig}

\newcommand{\plus}{\mbox{$^{+}$}}
\newcommand{\minus}{\mbox{$^{-}$}}

\newcommand{\electron}{\mbox{$e^{-}$}}			
\newcommand{\eminus}{\electron}
\newcommand{\positron}{\mbox{$e^{+}$}}
\newcommand{\eplus}{\positron}

\newcommand{\ee}{\eplus\eminus}			
 
\newcommand{\muplus}{\mbox{$\mu^{+}$}}		
\newcommand{\muminus}{\mbox{$\mu^{-}$}}

\newcommand{\mumu}{\muplus\muminus}

\newcommand{\tauplus}{\mbox{$\tau^{+}$}}
\newcommand{\tauminus}{\mbox{$\tau^{-}$}}

\newcommand{\tautau}{\tauplus\tauminus}
 
\newcommand{\bbar}{\mbox{$\bar{b}$} }

\newcommand{\pipi}{\mbox{$\pi$}^{+}\mbox{$\pi$}^{-}}

\newcommand{\bbbar}{\mbox{$b\bbar$}}

\newcommand{\goesto}{\mbox{$\rightarrow$}}

\newcommand{\chib}{\mbox{$\chi_{b}(1P)$}}
\newcommand{\chibp}{\mbox{$\chi_{b}(2P)$}}

\newcommand{\upsi}{\mbox{$\Upsilon$}{\rm (1S)}}
\newcommand{\upsii}{\mbox{$\Upsilon$}{\rm (2S)}}
\newcommand{\upsiii}{\mbox{$\Upsilon$}{\rm (3S)}}
\newcommand{\upsns}{\mbox{$\Upsilon$}{\rm (nS)}}
\newcommand{\upsms}{\mbox{$\Upsilon$}{\rm (mS)}}

\def \ba88{{\it Particles and Fields 3} (Proceedings of the 1988 Banff Summer
Institute on Particles and Fields), edited by A. N. Kamal and F. C. Khanna
(World Scientific, Singapore, 1989)}

\def \be87{{\it Proceedings of the Workshop on High Sensitivity Beauty
Physics at Fermilab,} Fermilab, Nov. 11-14, 1987, edited by A. J. Slaughter,
N. Lockyer, and M. Schmidt (Fermilab, Batavia, IL, 1988)}

\def \cp89{{\it CP Violation,} edited by C. Jarlskog (World Scientific,
Singapore, 1989)}
\def \dpf91{{\it The Vancouver Meeting - Particles and Fields '91}
(Division of Particles and Fields Meeting, American Physical Society,
Vancouver, Canada, Aug.~18-22, 1991), ed. by D. Axen, D. Bryman, and M. Comyn
(World Scientific, Singapore, 1992)}

\def \etal{{\it et\,al.}}

\def \ite{{\it et al.}}

\def \ky85{{\it Proceedings of the International Symposium on Lepton and
Photon Interactions at High Energy,} Kyoto, Aug.~19-24, 1985, edited by M.
Konuma and K. Takahashi (Kyoto Univ., Kyoto, 1985)}
\def \lat90{{\it Results and Perspectives in Particle Physics} (Proceedings of
Les Rencontres de Physique de la Vallee d'Aoste [4th], La Thuile, Italy, Mar.
18-24, 1990), edited by M. Greco (Editions Fronti\`eres, Gif-Sur-Yvette,
France,
1991)}

\def \plphgen#1
{{\it Proc. Int. Lepton-Photon Symposium (1991)} 
Geneva,
Switzerland, edited by S. Hagarty \etal (World Scientific, Singapore, 1991), 
p.#1}
\def \plphith#1
{{\it Proc. Int. Lepton-Photon Symposium (1993)} 
Ithaca, NY, edited by P. Drell and D. Rubin (AIP, New York, 1994), p.#1}

\def \lkl87{{\it Selected Topics in Electroweak Interactions} (Proceedings of
the Second Lake Louise Institute on New Frontiers in Particle Physics, 15 --
21 February, 1987), edited by J. M. Cameron \ite~(World Scientific, Singapore,
1987)}

\def \oxf65{{\it Proceedings of the Oxford International Conference on
Elementary Particles} 19/25 Sept.~1965, ed.~by T. R. Walsh (Chilton, Rutherford
High Energy Laboratory, 1966)}

\def \plb#1#2#3{{Phys. Lett. B} {\bf #1} #2 (#3)}

\def \prd#1#2#3{{Phys. Rev. D} {\bf#1}, #2 (#3)}
\def \prl#1#2#3{{Phys. Rev. Lett.} {\bf#1}, #2 (#3)}

\def \si90{25th International Conference on High Energy Physics, Singapore,
Aug. 2-8, 1990, Proceedings edited by K. K. Phua and Y. Yamaguchi (World
Scientific, Teaneck, N. J., 1991)}
 
\def \slac75{{\it Proceedings of the 1975 International Symposium on
Lepton and Photon Interactions at High Energies,} Stanford University, Aug.
21-27, 1975, edited by W. T. Kirk (SLAC, Stanford, CA, 1975)}
\def \slc87{{\it Proceedings of the Salt Lake City Meeting} (Division of
Particles and Fields, American Physical Society, Salt Lake City, Utah, 1987),
ed. by C. DeTar and J. S. Ball (World Scientific, Singapore, 1987)}
\def \smass82{{\it Proceedings of the 1982 DPF Summer Study on Elementary
Particle Physics and Future Facilities}, Snowmass, Colorado, edited by R.
Donaldson, R. Gustafson, and F. Paige (World Scientific, Singapore, 1982)}
\def \smass90{{\it Research Directions for the Decade} (Proceedings of the
1990 DPF Snowmass Workshop), edited by E. L. Berger (World Scientific,
Singapore, 1991)}

\def \tasi90{{\it Testing the Standard Model} (Proceedings of the 1990
Theoretical Advanced Study Institute in Elementary Particle Physics),
edited by M. Cveti\v{c} and P. Langacker (World Scientific, Singapore, 1991)}

\def\beq{\begin{equation}}
\def\eeq#1{\label{#1}\end{equation}}
\def\eeqn{\end{equation}}

\def\beqa{\begin{eqnarray}}
\def\eeqa#1{\label{#1}\end{eqnarray}}
\def\eeqan{\end{eqnarray}}

\let\bar=\overbar

\def\etal{{\it et al.}}

\def\M{{\cal M}}

\def\Dslash{\not{\hbox{\kern-4pt $D$}}}
\def\dslash{\not{\hbox{\kern-2pt $\del$}}}

\def\ee{e^+e^-}

\def\msb{{\bar{\ssstyle M \kern -1pt S}}}

\def\Title#1{\begin{center} {\Large {\bf #1} } \end{center}}

\begin{document}

\Title{Recent Results in Bottomonium Spectroscopy}

\begin{center}{\large \bf Contribution to the proceedings of HQL06,\\
Munich, October 16th-20th 2006}\end{center}

\bigskip\bigskip


\begin{raggedright}  

{\it Todd K. Pedlar\index{Pedlar, T. K.}\\
Department of Physics\\
Luther College\\
Decorah, IA, 52101 USA}
\bigskip\bigskip
\end{raggedright}

\section{Introduction}

Nearly thirty years ago, the E288 experiment at Fermilab reported~\cite{e288rept} the observation of a
significant excess in the number of events near 9.5 GeV in 
the spectrum of the invariant mass of $\mumu$ produced in a proton beam on a nuclear 
target.  Over the course of the subsequent two decades the spectrum of this newly-discovered
heavy quarkonium system was fleshed out in some detail.  Experiments which contributed to
this initial survey of the system included those whose methods of production
span the full spectrum of possible techniques, including the interaction of extracted hadron beams
on heavy targets, electron-positron annihilation and two-photon double-bremsstrahlung from 
very high energy electron and positron beams.  Compared to charmonium, the bottomonium spectrum
below open flavor threshold is richer, offering more opportunity to flesh out the details of the
interactions between heavy quarks.  Despite the two decades of research on the spectrum of this beautiful
system, many things remain hidden and left to be discovered.  We describe the status of several
studies in bottonium spectroscopy made over the course of the past several years at CLEO,
BaBar and Belle.

As evidenced by the prodigious output of the Quarkonium Working Group,~\cite{qwg} heavy quarkonia offer an 
important experimental laboratory for the understanding of quark-antiquark interactions.  The bottomonium system,
free as it is of some of the more severe relativistic complications due to the very large mass of the bottom quark, is a particularly advantageous system to study.  Among the important quantities that one would like to measure 
precisely in the bottomonium spectrum are the masses, hyperfine splittings, dilepton decay widths, etc.  These are key inputs and checks for Lattice QCD calculations, which are steadily improving in precision.  Such data, and additionally measurements of the details of the P-wave fine structure can also aid the development of QCD and potential models.  Hadronic decay rates and detailed studies of angular distributions and invariant mass spectra are important in developing our theoretical understanding of hadronization.  

In each of these cases, comparison to similarly precise measurements in the charmonium system that have been performed by CLEO, E835 and other recent experiments, provide a very nice foundation for improving theoretical models that describe the dynamics of heavy quarks.

\section{The Experimental Situation Circa 2001}

\begin{figure}[tb]
\begin{center}
\epsfig{file=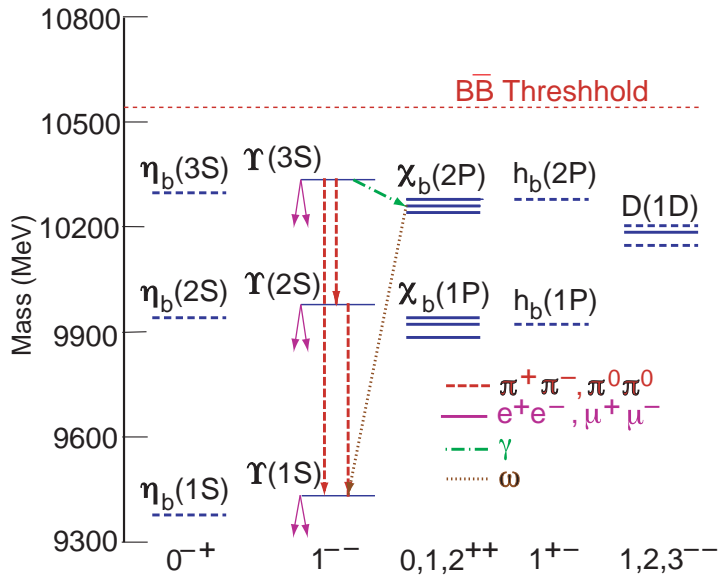,width=4.5in}
\caption{Schematic showing the full $\bbbar$ spectrum below open beauty threshold.}
\label{bbspec}
\end{center}
\end{figure}
At the beginning of 2001, the spectrum of bottomonium was known, in its gross features, relatively well (see Figure~\ref{bbspec}). Three
triplet-S states, the $\upsiii$, $\upsii$ and $\upsi$ had been observed, with known branching ratios to lepton pairs; two pairs of triplet-P states ($\chib$ and $\chibp$) had all been observed in radiative transitions from the higher $\upsns$ states, and many had been observed to decay radiatively to lower $\upsms$ states.  The masses of all nine of these states were known to varying degrees of precision.  Leptonic branching ratios and/or partial widths were not all well known.

From 2001 to 2002, CLEO took data at center of mass energies 
on or near the three lower-lying triplet-S bottomonium resonances, expending from 1.2 to 1.5 $fb^{-1}$ in 
each case.  The number of $\upsns$ decays observed were approximately 22, 9 and 6 million, for $\upsi$,
$\upsii$ and $\upsi$, respectively.  These data samples were at the time the largest single 
samples of each state, representing factors of ten- to twenty-fold statistical improvement.  

\section{New Leptonic Decay Measurements from $e$ to $\tau$}

\begin{figure}[tb]
\begin{center}
\epsfig{file=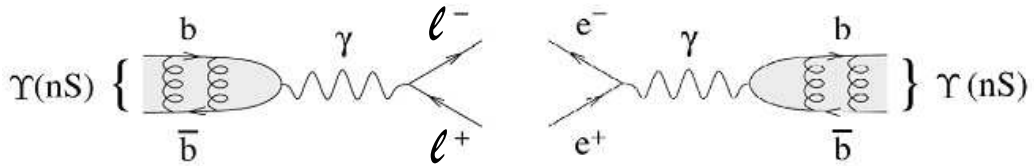,width=5.5in}
\caption{Schematic showing dilepton decay and production vertices involving $\upsns$.}
\label{dilepvtx}
\end{center}
\end{figure}

The leptonic partial width of a quarkonium state is one of the basic parameters calculable in LQCD, and is 
therefore a valuable parameter to be explored with precision.  In Figure~\ref{dilepvtx} are shown Feynman diagrams illustrating the heavy quarkonium production and decay vertices involving a lepton pair. At CLEO, we have recently measured leptonic widths or branching ratios for all three $\upsns$ states that lie below open-bottom threshold to all three lepton flavors ($\ee$,~\cite{cleoee} $\mumu$,~\cite{cleomumu} $\tau\plus\tau\minus$~\cite{cleotautau}).  We describe these measurements in chronological order. 
 
\subsection{$\mumu$ Branching Ratios}

The branching ratios of $\upsns$ to $\mumu$ involve a relatively simple measurement: the observation of a final state consisting only of a high-momentum muon pair.  Dominant systematic uncertainties for these measurements are the hadronic trigger efficiency, muon detection efficiency and the scale factor which is used in the background subtraction.  Overall systematic uncertainties are 2.7, 3.7 and 4.1$\%$, for the branching ratio measurements of $\upsi$, $\upsii$ and $\upsiii$, respectively.  

The new measurements by CLEO of these branching ratios represent, in the case of $\upsii$ and $\upsiii$ in particular, substantial improvements in precision - and are significantly different than the previously-reported world average values.  Because of this, these measurements have significant impact on 
many of the branching ratios for cascade decays that end in $\upsii$. They also impact the total widths of 
$\upsii$ and $\upsiii$.  Using the
PDG2004 values for the quantity $\Gamma_{ee}\Gamma_{had}/\Gamma_{tot}$, we obtain
$\Gamma(1S)=(52.8\pm 1.8)$keV, $\Gamma(2S)=(29.0\pm 1.6)$ keV and $\Gamma(3S)=(20.3\pm 2.1)$keV.  This should be compared with the previous~\cite{pdg04} evaluations in the 2004 version of the PDG of $\Gamma(1S)=(53.0\pm 1.5)$keV, $\Gamma(2S)=(43\pm 6)$ keV and $\Gamma(3S)=(26.3\pm 3.4)$keV.

\begin{figure}[tpb]
\begin{center}
\epsfig{file=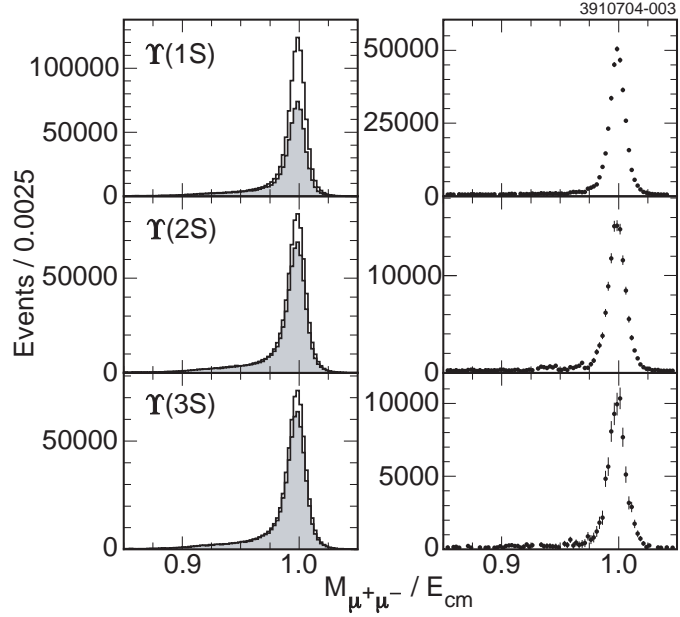,width=3.5in}
\caption{Reconstructed $\mumu$ as a function of $\mumu$ invariant mass for each of the three
low-lying $\upsns$ states.  (left) Open histogram: reconstructed events; shaded histogram: scaled off-resonance
data for continuum subtraction. (right) Difference between the two histograms at left reveals the decay of $\upsns$ to $\mumu$.}
\label{muresults}
\end{center}
\end{figure}


\begin{table}[bp]
\begin{center}
\begin{tabular}{l|ll}
\hline
Quantity & New CLEO Result & PDG2004 \\
\hline
${\cal{B}}_{\mu\mu}(\upsi)$ 		& $(2.49\pm 0.02\pm 0.07)\%$ 	&$(2.48\pm 0.06)\%$\\
${\cal{B}}_{\mu\mu}(\upsii)$ 	& $(2.03\pm 0.03\pm 0.08)\%$	&$(1.31\pm 0.21)\%$\\
${\cal{B}}_{\mu\mu}(\upsiii)$	&	$(2.39\pm 0.07\pm 0.10)\%$	&$(1.81\pm 0.17)\%$\\
\hline
\end{tabular}
\label{mumufrac}
\caption{New measurements of $B(\upsns\goesto\mumu)$ from CLEO.~\cite{cleomumu}}
\end{center}
\end{table}
\subsection{$\ee$ Partial Widths}

\begin{figure}[t]
\begin{center}
\epsfig{file=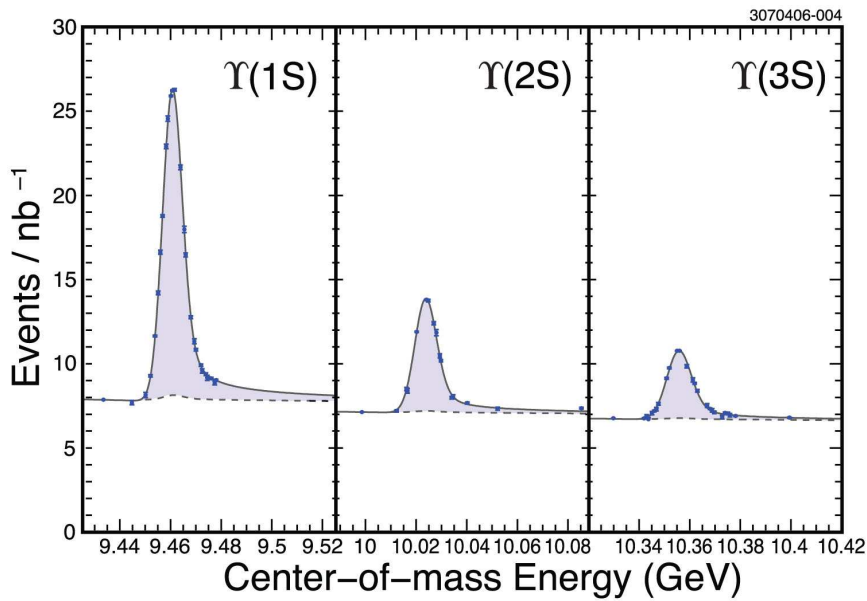,width=4.5in}
\caption{Hadronic yield measured by CLEO during scans of the center of mass regions near the $\upsns$ masses. The points represent the data, the solid line is the fit, and the dashed line the sum of backgrounds, which are dominantly hadronic continuum and radiative Bhabhas.  The shaded area represents the yield from which $\Gamma_{ee}$ is obtained.~\cite{cleoee}}
\label{eeresults}
\end{center}
\end{figure}

Because at CLEO $\upsns$ states are produced at resonance by annihilation of $\eplus$ and $\eminus $ beams, the partial width of these states to $\ee$ are not measured by studying the $\ee$ final state, but by observing the decay of these states to hadrons at various center of mass energies $(\sqrt{s})$ that span the width of the state and integrating under the excitation curve thus produced.  In these measurements a systematic uncertainty of $1.5-1.8\%$ was achieved - the chief contributing uncertainty being the $1.3\%$ uncertainty from the overall luminosity scale.  From the fit to the yields (see Figure~\ref{eeresults}) is obtained the partial width product $\Gamma_{ee}\Gamma_{had}/\Gamma_{tot}$.  Applying an assumption of lepton universality, and using the recent measurement of $B(\mumu)$ for all three states by CLEO,~\cite{cleomumu} this product is then converted into the final results for $\Gamma_{ee}$, thus:

\[\Gamma_{ee}=\frac{\Gamma{ee}\Gamma_{had}/\Gamma_{tot}}{1-3B(\mumu)}.\]

\begin{table}[pt]
\begin{center}
\begin{tabular}{l|l}
\hline
$\Gamma_{ee}\Gamma_{had}/\Gamma_{tot}(\upsi)$ : & $1.252\pm 0.004 \pm 0.019$ keV\\
$\Gamma_{ee}\Gamma_{had}/\Gamma_{tot}(\upsii)$ : & $0.581\pm 0.004\pm 0.009$ keV\\
$\Gamma_{ee}\Gamma_{had}/\Gamma_{tot}(\upsiii)$: & $0.413\pm 0.004\pm 0.006$ keV\\
\hline
$\Gamma_{ee}(\upsi)$ : & $1.354\pm 0.004\pm 0.020$ keV\\
$\Gamma_{ee}(\upsii)$ : & $0.619\pm 0.004\pm 0.010$  keV\\
$\Gamma_{ee}(\upsiii)$: & $0.446\pm 0.004\pm 0.007$ keV\\
\hline
$\Gamma_{ee}(\upsii)/\Gamma_{ee}(\upsi)$ : & $0.457\pm 0.004\pm 0.004$\\
$\Gamma_{ee}(\upsiii)/\Gamma_{ee}(\upsi)$ : & $0.329\pm 0.003\pm 0.003$\\
$\Gamma_{ee}(\upsiii)/\Gamma_{ee}(\upsii)$: & $0.720\pm 0.009\pm 0.007$\\
\hline\end{tabular}
\caption{Various combinations of partial and total widths for the three
triplet-S bottomonium resonances.  The top three rows contain the primary measurements
made by CLEO~\cite{cleoee}.  The second two sets of three rows are derived quantities 
using combinations of measurements and/or other experimental inputs~\cite{cleomumu,pdg04}.}
\label{eeratios}
\end{center}
\end{table}
\begin{figure}[pb]
\begin{center}
\epsfig{file=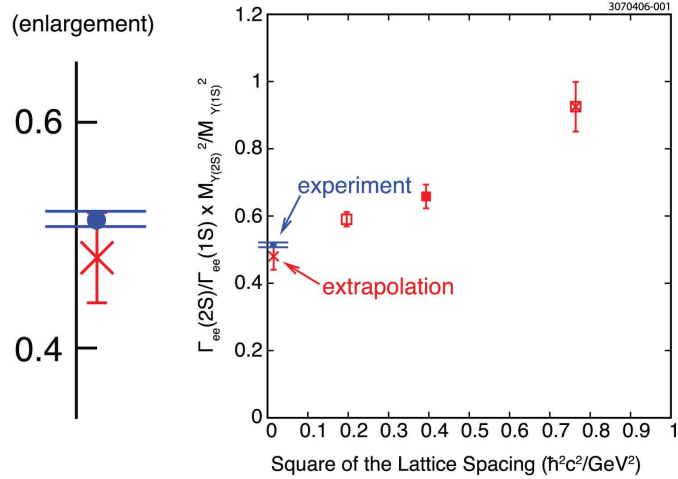,width=3.5in}
\caption{Comparison between the CLEO measurement and the unquenched
lattice QCD calculation of the quantity 
$\Gamma_{ee}(2S)M^2(2S))/(\Gamma_{ee}(1S)M^2(1S))$. 
~\cite{cleoee,lgt}}
\label{lgtcompare}
\end{center}
\end{figure}
The results are tabulated in Table~\ref{eeratios}.  The best object for comparison to lattice calculations is the ratio of $\ee$ partial widths times the square of the mass of the state for pairs of states.  For instance, we compare in Figure~\ref{lgtcompare} the measured ratio $\Gamma_{ee}(2S)M^2(2S))/(\Gamma_{ee}(1S)M^2(1S))$.  The most recent lattice result for this quantity compares well with the experimental result, with theoretical uncertainties somewhat larger than the uncertainties of the experimental result:
\begin{eqnarray*}
\Gamma_{ee}(2S)M^2(2S))/(\Gamma_{ee}(1S)M^2(1S))&=& 0.514\pm 0.007\mbox{ (CLEO, \cite{cleoee})}\\
&=& 0.48\pm 0.05\mbox{ (LGT,~\cite{lgt}).}
\end{eqnarray*}

\subsection{$\tau\plus\tau\minus$ Branching Ratios}

Prior to the CLEO result which is presented here, only the $\upsi$ decay to $\tautau$ was at all well measured.
$\upsii$ had been observed to decay to $\tautau$, but $\upsiii$ had never been observed in this decay mode.
For this study, each $\tau$ is observed to decay via one of its one-prong decay modes.  One of the benefits of this selection is a very clean final state involving just two tracks, but one which is easily distinguishable from Bhabha, $\upsns\goesto\ee$ or $\upsns\goesto\mumu$ by virtue of the large missing energy in the event that is taken up by unobserved neutrinos.  Furthermore, the results of this analysis are quoted as a ratio to the very well measured $\mumu$ branching ratios presented above, so that many systematic uncertainties are either cancelled or very greatly reduced.  
\begin{figure}[t]
\begin{center}
\epsfig{file=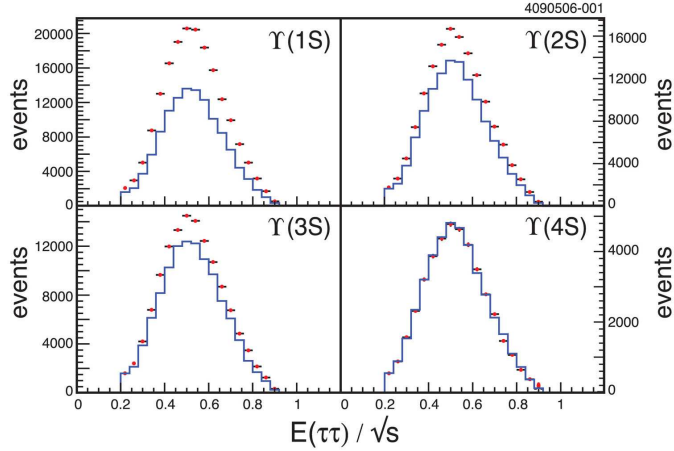,width=3.5in}
\caption{Ratio of observed $\tau$ pair energy to $\sqrt{s}$ for 
each of the four $\upsns$ resonances.  The points represent the data
and the solid histogram the scaled continuum.  From the excess above scaled continuum,
we obtain the final event yield, and subsequently $B(\tautau)$.  The plot of data
taken at the $\Upsilon(4S)$, which shows
agreement between the scaled continuum and the on-resonance data is
included to demonstrate the validity of the background subtraction.~\cite{cleotautau}}
\label{tauresults}
\end{center}
\end{figure}
\begin{table}[t]
\begin{center}
\begin{tabular}{l|ll}
\hline
& $B(\tautau)/B(\mumu)$ & $B(\tau\tau)(\%)$\\
\hline
$\upsi$  & $1.02\pm 0.02\pm 0.05$ & $2.54\pm 0.04\pm 0.12$\\
$\upsii$ & $1.04\pm 0.04\pm 0.05$ & $2.11\pm 0.07\pm 0.13$\\
$\upsiii$ & $1.07\pm 0.08\pm 0.05$ & $2.55\pm 0.19\pm 0.15$\\
\hline
\end{tabular}
\caption{Ratios of $B(\tau\plus\tau\minus)$ to $B(\mumu)$, and 
$B(\tautau)$, calculated using the most recent CLEO measurements of
$B(\mumu)$~\cite{cleomumu} for the three
triplet-S bottomonium resonances.~\cite{cleotautau}}
\label{tauratios}
\end{center}
\end{table}
These new measurements are of particular interest as tests of lepton universality
in the decays of $\upsns$ - they demonstrate consistency with Standard Model expectations,
and represent the most precise single measurement of $B(\upsi\goesto\tau\tau)$, a very
much improved value of $B(\upsii\goesto\tau\tau)$ and the first ever measurement
of $B(\upsiii\goesto\tau\tau)$.  
\section{New Studies of Hadronic Transitions}

\begin{figure}[hb]
\begin{center}
\epsfig{file=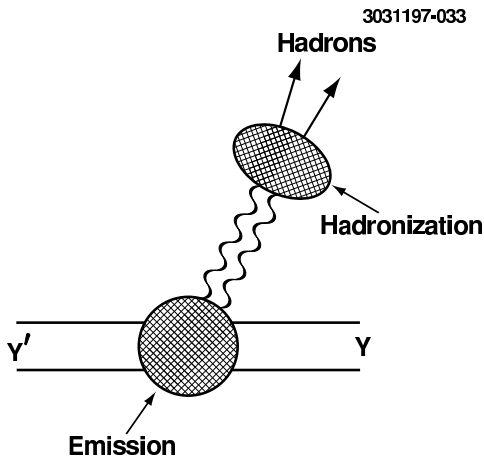,width=2.5in}
\caption{Schematic illustrating the transtion from one $\upsns$ state
to a lower state by emission of hadrons.}
\label{hadtrans}
\end{center}
\end{figure}

Measurements of the decay rates and kinematic characteristics 
of hadronic transitions within the bottomonium system, 
which are generally understood theoretically in terms of 
a multipole expansion model~\cite{gottfried}, (see Figure~\ref{hadtrans}) are important tools
for improving our understanding of the hadronization process.  Ever since the discovery of
the bottomonium dipion transitions by CLEO and others, there has been a well-known difficulty 
in the describing the various transitions in a single theoretical framework.  In particular,
the $\upsiii\goesto\upsi\pi\pi$ decays are of interest. 
 
In the past year, both Belle and BaBar have produced interesting results involving hadronic transitions 
from the $\Upsilon(4S)$. 
Together with recent results from CLEO on hadronic transitions between $\upsiii$ and $\upsii$ and the lower $\upsms$ resonances, an intriguing picture is emerging concerning the distributions of dipion invariant masses.  

\subsection{Studies of Hadronic Transitions at Belle}
The Belle study of $\Upsilon(4S)\goesto\pipi\Upsilon(1S)$,~\cite{belle4s} was based on a sample of integrated luminosity $398\mbox{ fb}^{-1}$ (386 million $\Upsilon(4S)$ decays).  In this study, the daughter $\upsi$ was observed via its decay to $\mumu$.  After requiring a $\mumu$ candidate with an invariant mass in the vicinity of $M(\upsi) = 9.460$ GeV,  an additional pair of pions of opposite charge was required, with the criterion 
$\Delta M\equiv \M(\mumu\pipi) - M(\mumu)$ satisfying the expectation that it be consistent with  
 $10.580-9.460\mbox{ GeV } = 1.120\mbox{ GeV}$.  A significant source of background for this analysis is 
 the process $\ee\goesto\gamma\mu\mu$, in which the $\gamma$ converts in detector material and fakes the
 $\pipi$.

\begin{figure}[tb]
\begin{center}
\epsfig{file=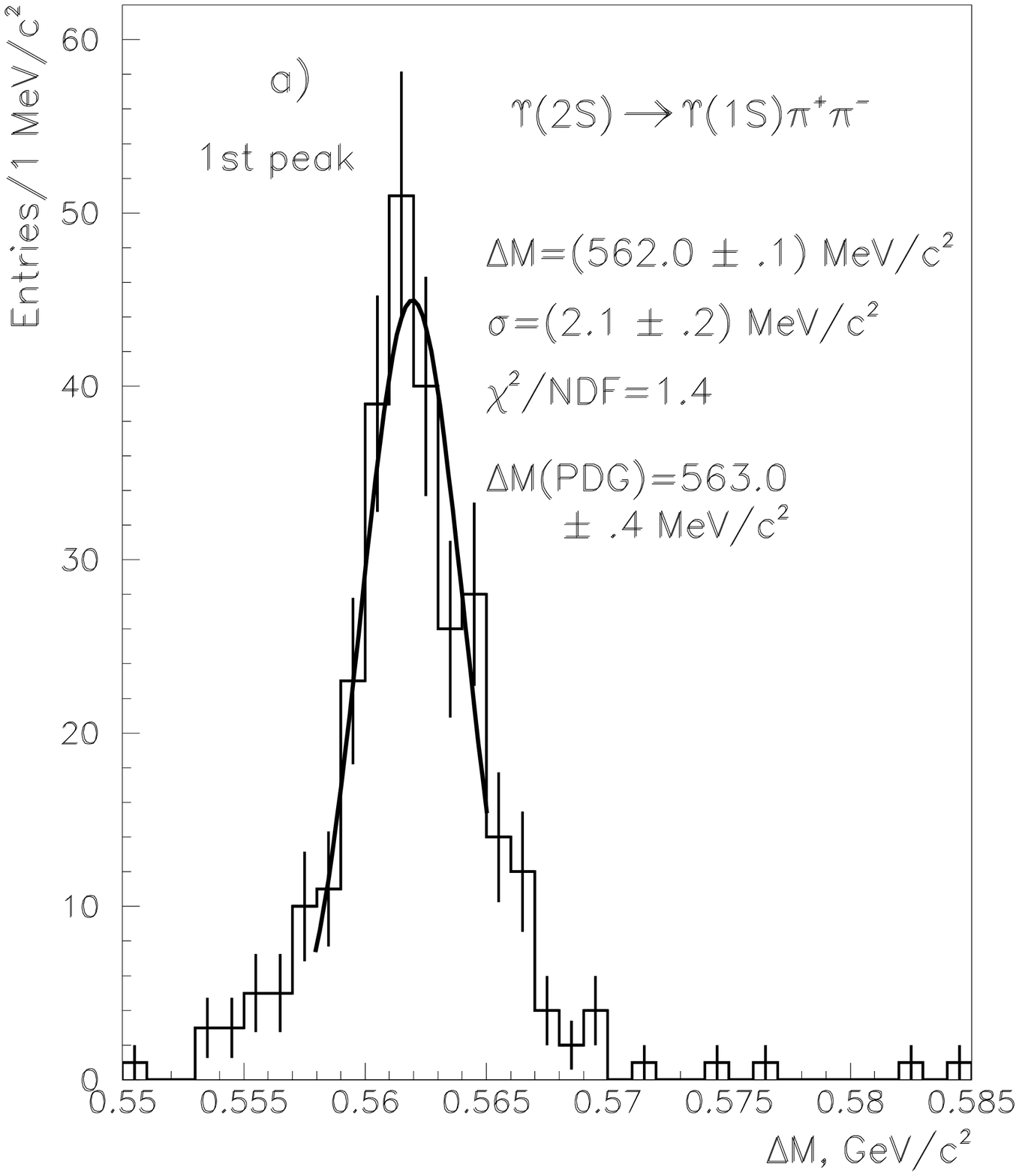, width=1.9in}
\epsfig{file=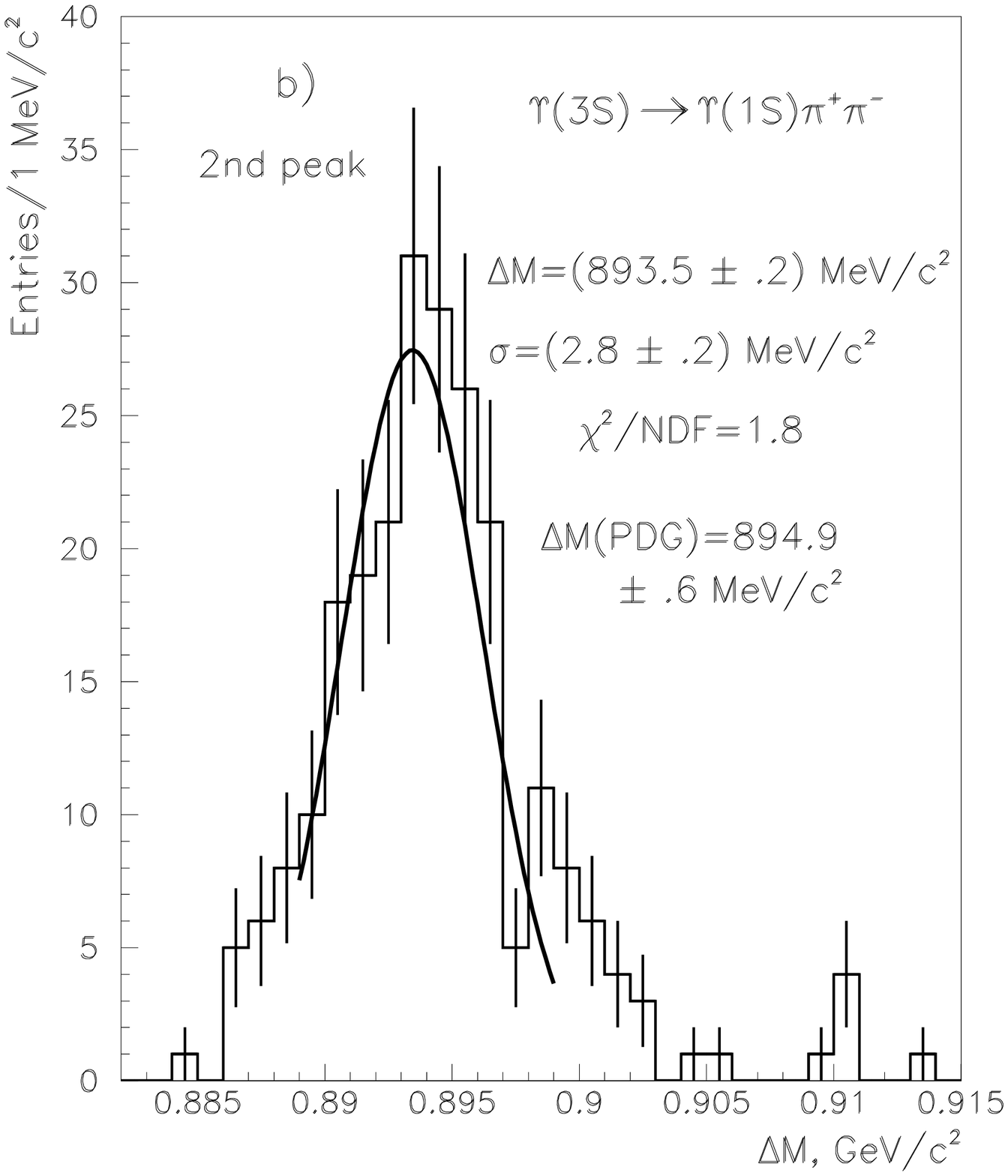, width=1.9in}
\epsfig{file=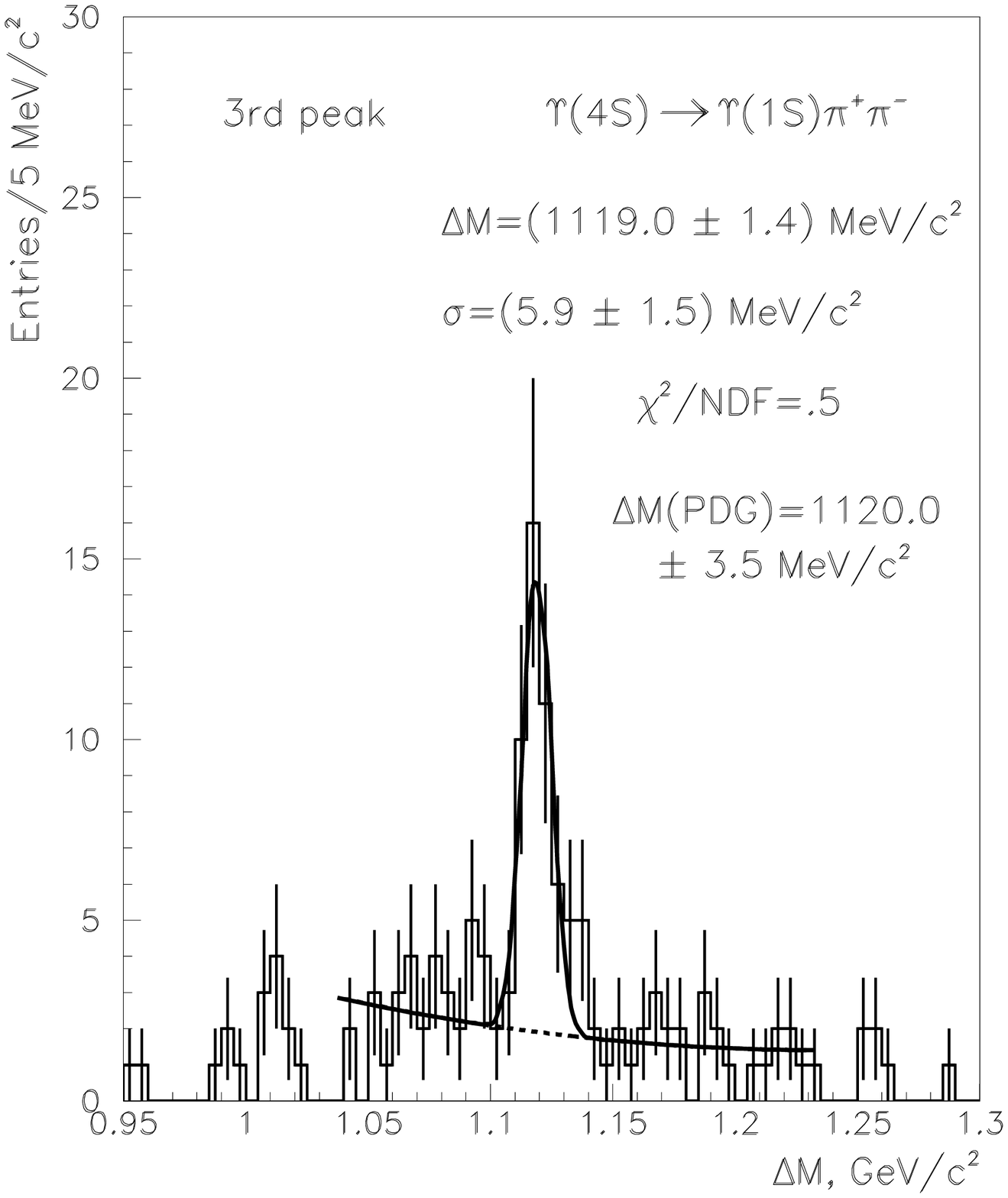, width=1.9in}
\caption{Plots from Belle of the quantity $\Delta M$, the mass difference between the 
combination of $\mumu\pipi$ and the $\mumu$ alone, in the region expected for
the observation of (left) $\upsii\goesto\pipi\upsi$, (center) $\upsiii\goesto\pipi\upsi$ and
(right) $\Upsilon(4S)\goesto\pipi\upsi$.~\cite{belle4s}}
\label{bellefig}
\end{center}
\end{figure}
In order to obtain a yield of the desired signal events, the $\Delta M$ spectrum is fitted.  Shown in
 Figure~\ref{bellefig} are the signals near the appropriate $\Delta M$ regions for the observation of
 processes involving either the resonant decay of $\Upsilon(4S)$ or the radiative return process 
 $\ee\goesto\gamma\upsiii,\upsii$ and the subsequent decay $\upsiii,\upsii\goesto\pipi\upsi$.  After
 correcting for acceptance and efficiency, a branching ratio of 
\[B(\Upsilon(4S)\goesto\pipi\upsi) = (1.1\pm 0.2\pm 0.4)\times 10^{-4}\]
was obtained.  Using the PDG value for  the full width of $\Upsilon(4S)$, they obtain a partial width
\[\Gamma(\Upsilon(4S)\goesto\pipi\upsi) = (2.2\pm 1.0)\mbox{ keV.}\]
  This value can be compared to the decay
  partial widths to $\pipi\upsi$ for the $\upsiii$ and $\upsii$ of $8.1\pm 1.2$ keV and $1.2\pm 0.2$ keV, respectively.
 
\subsection{Studies of Hadronic Transitions at BaBar}

The BaBar study of $\Upsilon(4S)$ hadronic transitions was based on a sample of integrated luminosity 
$211\mbox{ fb}^{-1}$ (230 million $\Upsilon(4S)$ decays). 
Both $\Upsilon(4S)\goesto\pipi\upsi$ and
$\Upsilon(4S)\goesto\pipi\upsii$ transitions were observed using a similar analysis method to that used by Belle.~\cite{babar4s}  Their selection of the $\mumu\pipi$ final state required a pair
of charged particles consistent with being $\mumu$ and a second pair of tracks which are not 
consistent with being $\ee$, in order to remove the events in which the $\gamma$ produced in 
a radiative mu pair event converts to $\ee$ and would otherwise fake the $\pipi$ signal. 

\begin{figure}[pt]
\begin{center}
\epsfig{file=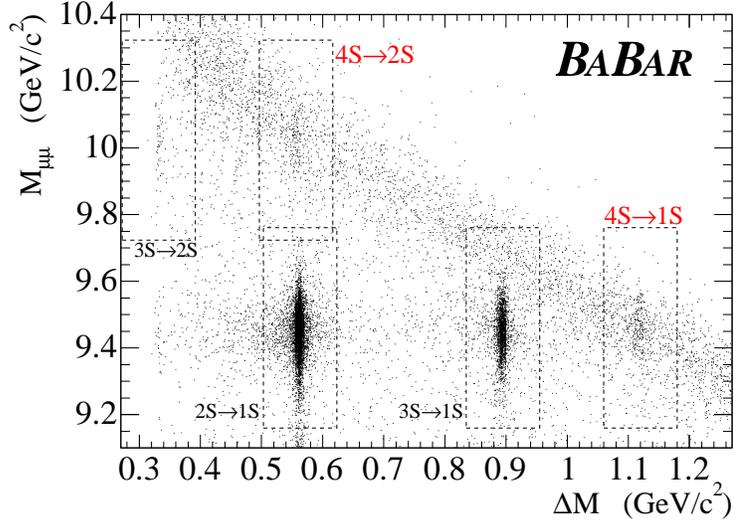, width=4.in}
\caption{Two-dimensional plot from BaBar of
$M(\mumu)$, the invariant mass of the $\mumu$ candidate vs. 
$\Delta M$, the mass difference between the combination of $\mumu\pipi$ and the $\mumu$ alone.~\cite{babar4s}}
\label{babar2dplot}
\end{center}
\end{figure}

\begin{figure}[pb]
\begin{center}
\epsfig{file=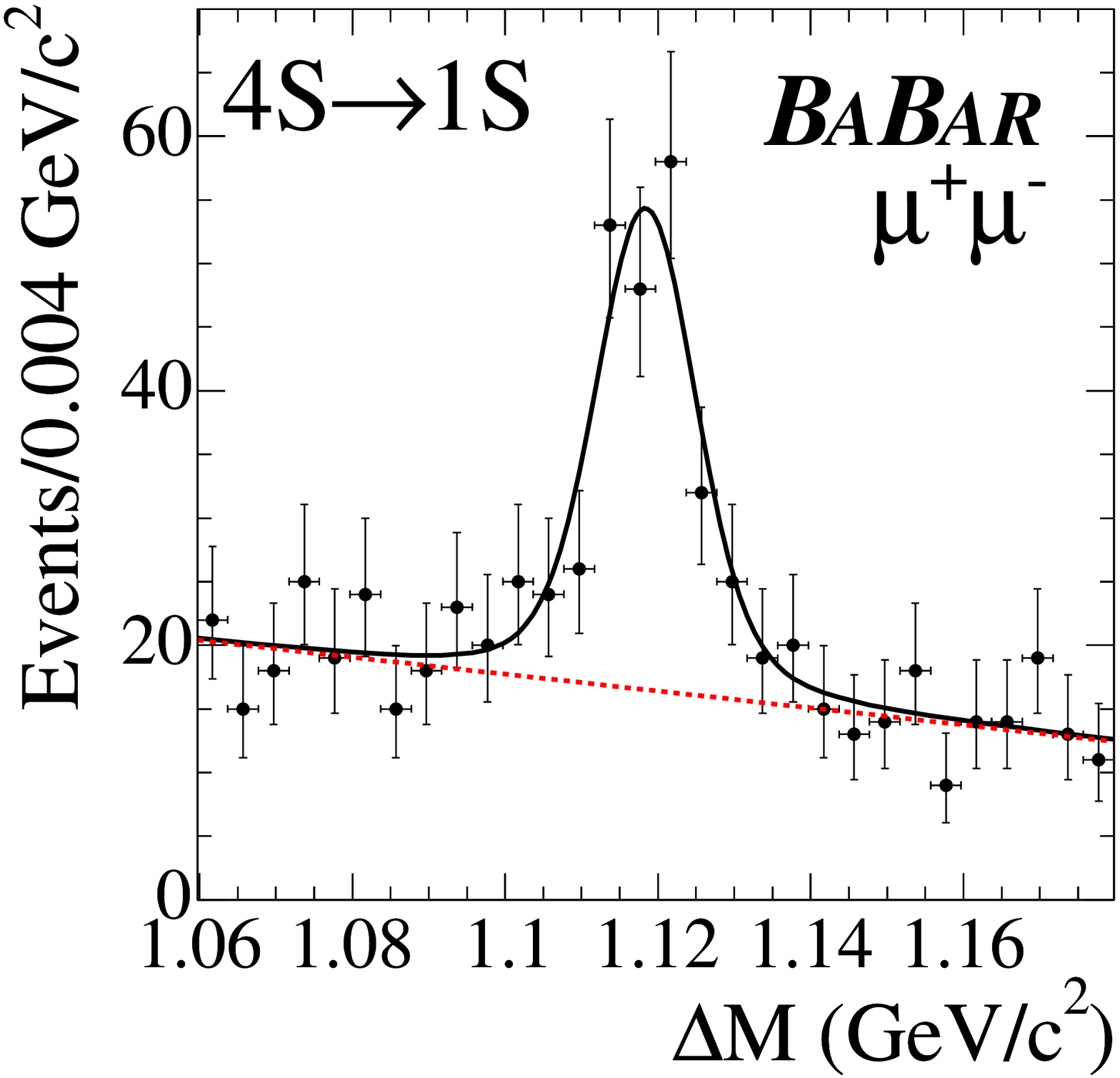, width=2.6in}
\epsfig{file=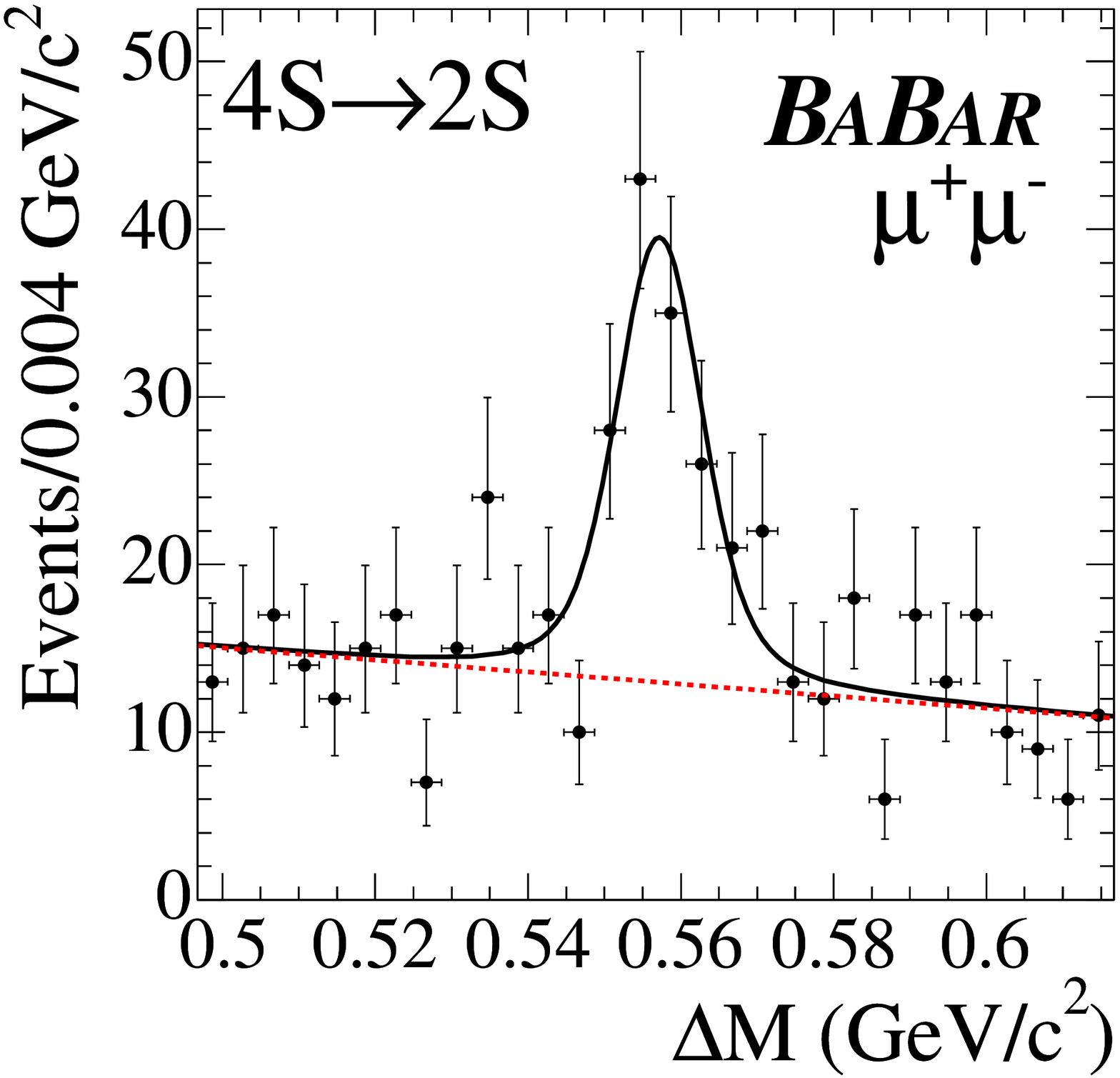, width=2.6in}
\caption{Plots from BaBar of the quantity $\Delta M$, the mass difference between the 
combination of $\mumu\pipi$ and the $\mumu$ alone, in the region expected for
the observation of (left) $\Upsilon(4S)\goesto\pipi\upsi$ and (right) $\Upsilon(4S)\goesto\pipi\upsii$.~\cite{babar4s}}
\label{babar4s}
\end{center}
\end{figure}

BaBar defines the same quantity $\Delta M$ as do their counterparts at Belle, and similarly fit
the $\Delta M$ spectra obtained by selecting appropriate regions in the two-dimensional plane
of $M(\mumu)$ vs. $\Delta M$ (see Figure~\ref{babar2dplot}).  The two $\Delta M$ spectra used
in the observation of $\Upsilon(4S)\goesto\pipi\upsi$ and $\Upsilon(4S)\goesto\pipi\upsii$ are
shown in Figure~\ref{babar4s}.  From the fits to the data shown are obtained yields (signal 
significances) of 
$167\pm 19 (10.3\sigma)$ and $97\pm 15 (7.3\sigma)$ for the transitions 
terminating in $\upsi$ and $\upsii$, respectively. (See Table~\ref{babartable})  
From these yields, the following
product branching fractions are obtained: 
\begin{eqnarray*}
B(\Upsilon(4S)\goesto\upsi\pipi)\times B_{\mumu}(\upsi) = (2.23\pm 0.25\pm 0.27)\times 10^{-6}\\
B(\Upsilon(4S)\goesto\upsii\pipi)\times B_{\mumu}(\upsii) = (1.69\pm 0.26\pm 0.20)\times 10^{-6}\\
\end{eqnarray*}

These products may then, using the most recent CLEO results for $B_{\mumu}(\upsns)$,~\cite{cleomumu} and a
recent measurement (BaBar) of $\Gamma_{tot}(\Upsilon(4S))$,~\cite{babar4swidth} be used to obtain the following
partial widths $\Gamma(\Upsilon(4S)\goesto\upsns\pipi$:
\begin{eqnarray*}
\Gamma(\Upsilon(4S)\goesto\upsi\pipi)= 1.8\pm 0.4 \mbox{ keV}\\
\Gamma(\Upsilon(4S)\goesto\upsii\pipi)= 1.7\pm 0.5 \mbox{ keV}\\
\end{eqnarray*}

\begin{table}[t]
\begin{center}
\begin{tabular}{l|l|ll}
\hline
Experiment & Transition & $B_{\pipi}$ &$\Gamma_{\pipi}$\\
&& $ (\times 10^{-4})$ & (keV)\\
\hline
Belle & $4S\goesto 1S$ & $1.13\pm 0.45$ & $2.3\pm 0.9$\\
BaBar & $4S\goesto 1S$ & $0.90\pm 0.15$& $1.8\pm 0.4$\\
BaBar & $4S\goesto 2S$ & $0.88\pm 0.19$ & $1.8\pm 0.5$\\
\hline
\end{tabular}
\end{center}
\caption{Results of measurements by Belle~\cite{belle4s}
 and from BaBar~\cite{babar4s}.  
For ease of comparison, the branching ratios and the partial widths have been
recalculated using common inputs: the PDG06 average value for 
$\Gamma(4S)$~\cite{pdg06} and the most recent CLEO results for $B(\upsns\goesto\mumu)$.~\cite{cleomumu}  
Belle has since the conference updated their
result, basing a new measurement on 477$fb^{-1}$ of integrated luminosity at $\Upsilon(4S)$, 
obtaining $B(\pipi)=1.77\pm 0.23 \times 10^{-4}$ and $\Gamma(\pipi)=3.7\pm 0.9$ keV.~\cite{newbelle}}
\label{babartable}
\end{table}

\subsection{Studies of Dipion Invariant Mass Spectra}

\begin{figure}[tb]
\begin{center}
\epsfig{file=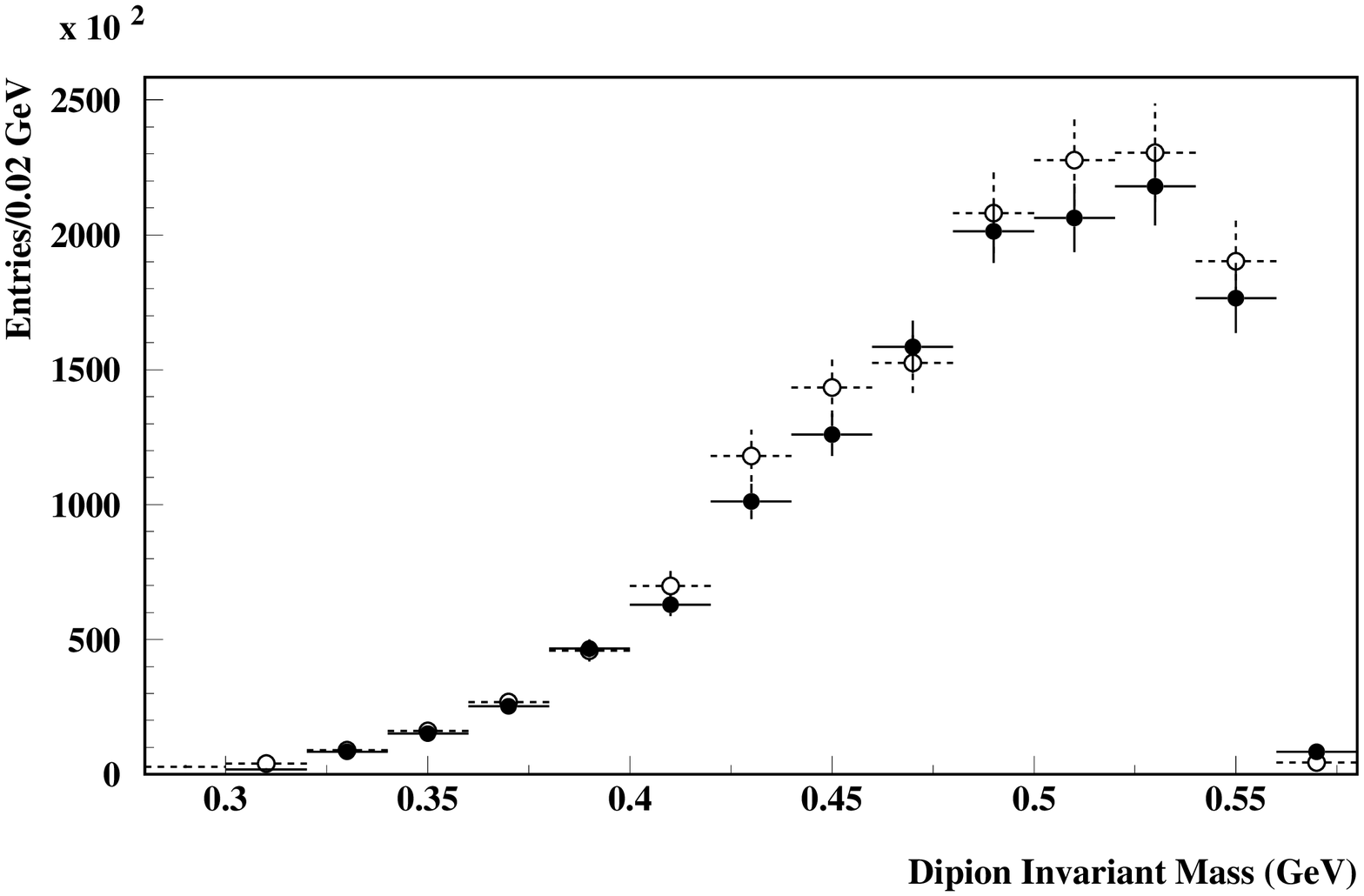,width=2.2in}
\epsfig{file=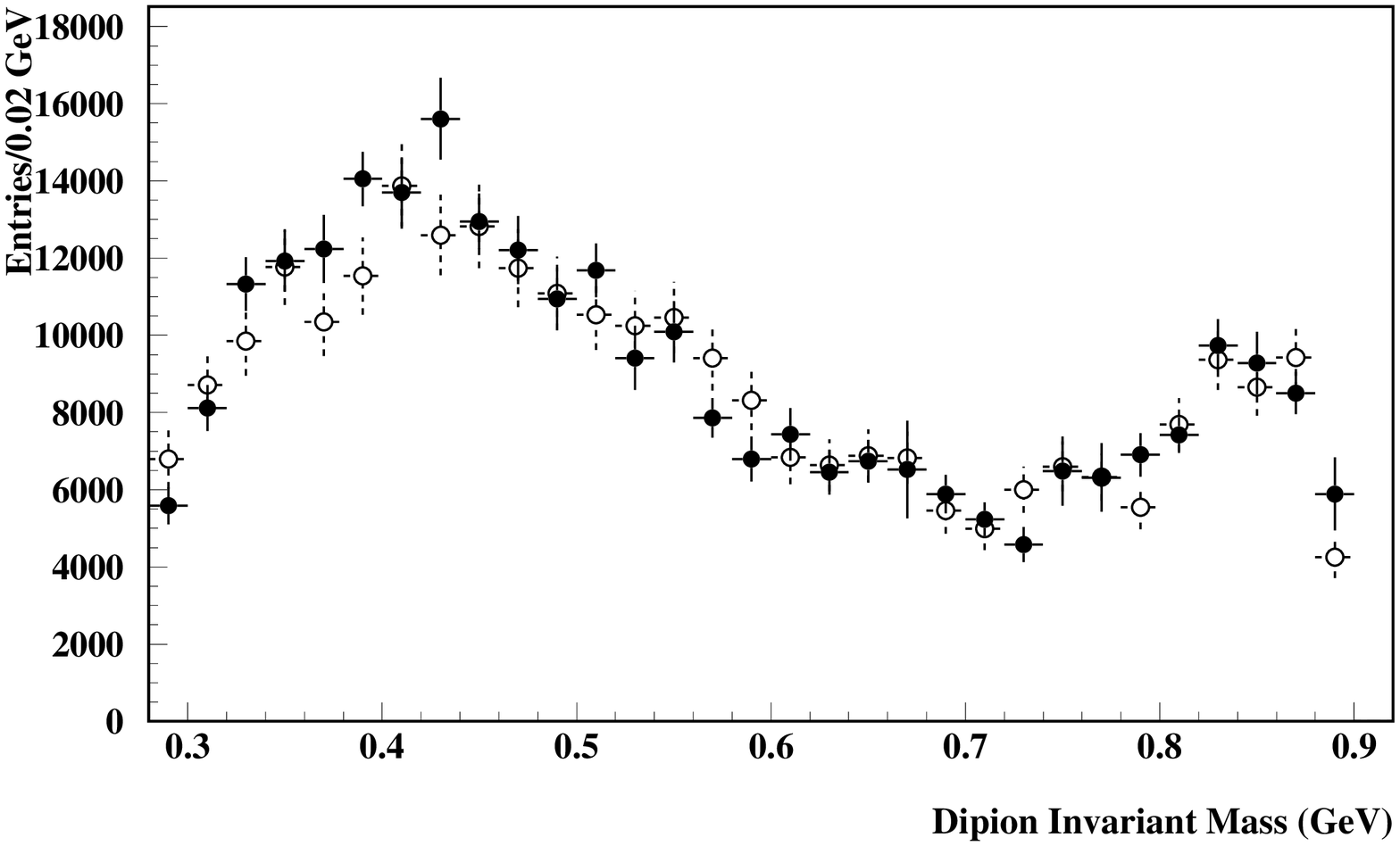,width=2.2in}
\epsfig{file=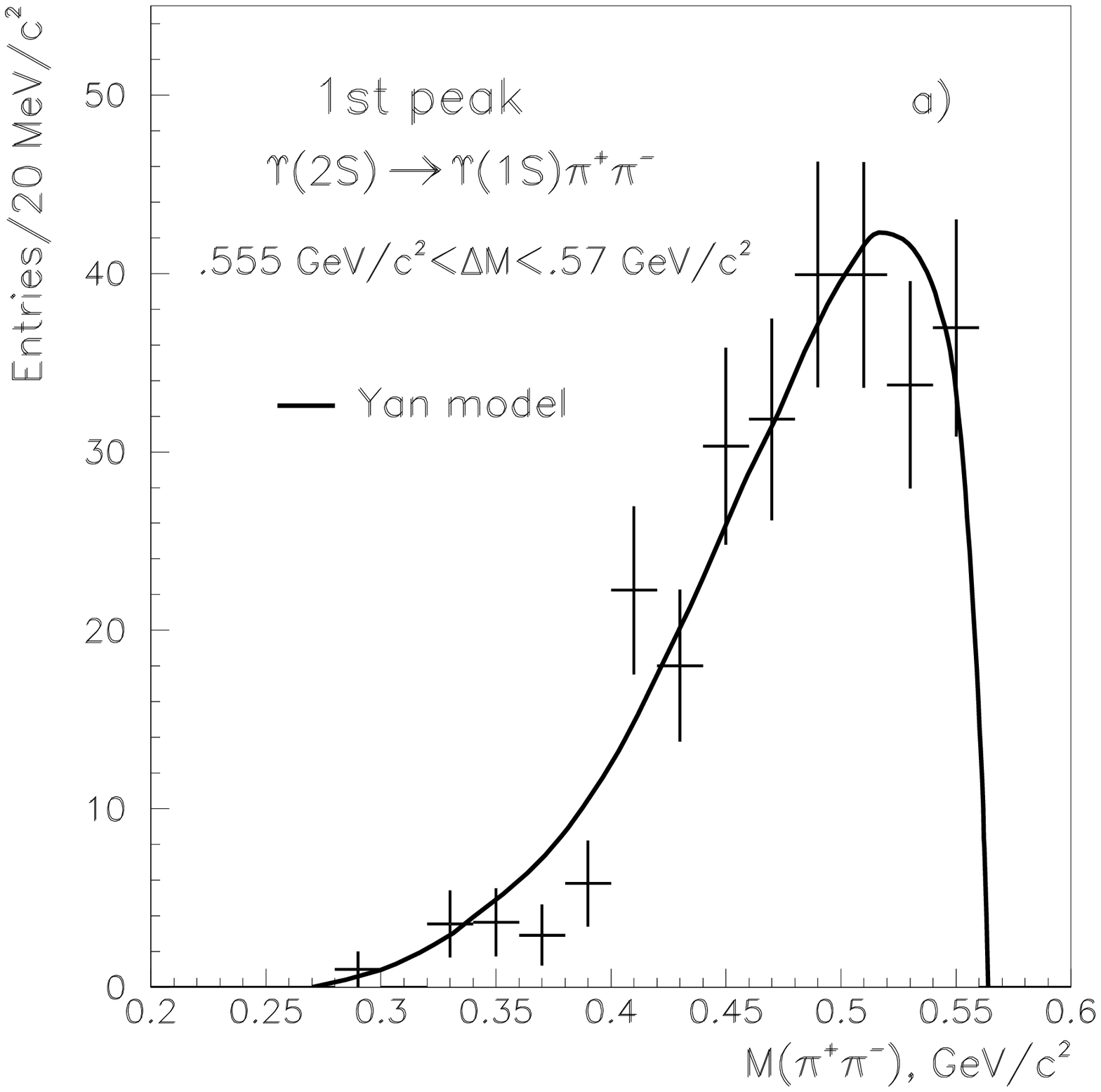,width=2.2in}
\epsfig{file=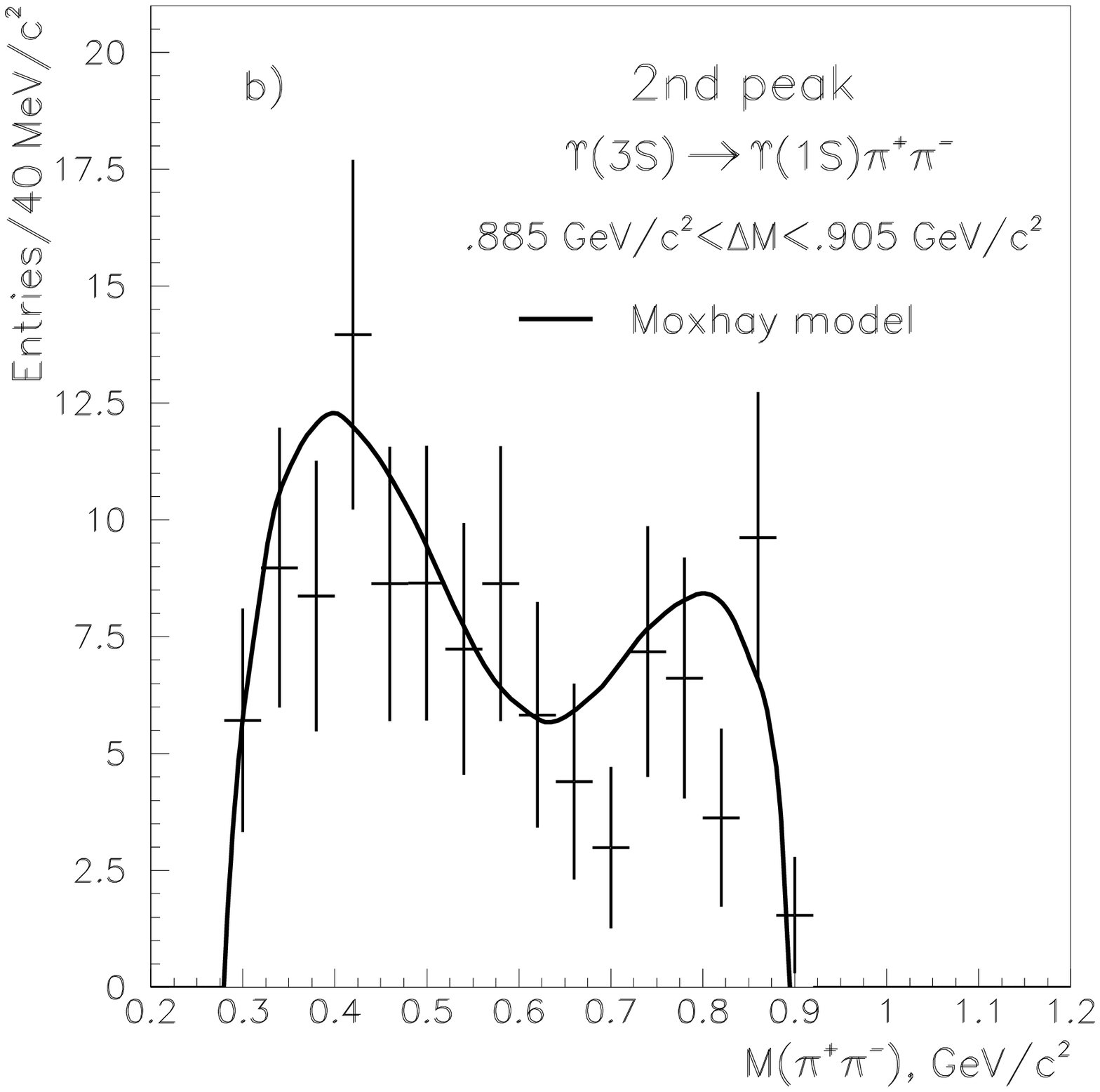,width=2.2in}
\epsfig{file=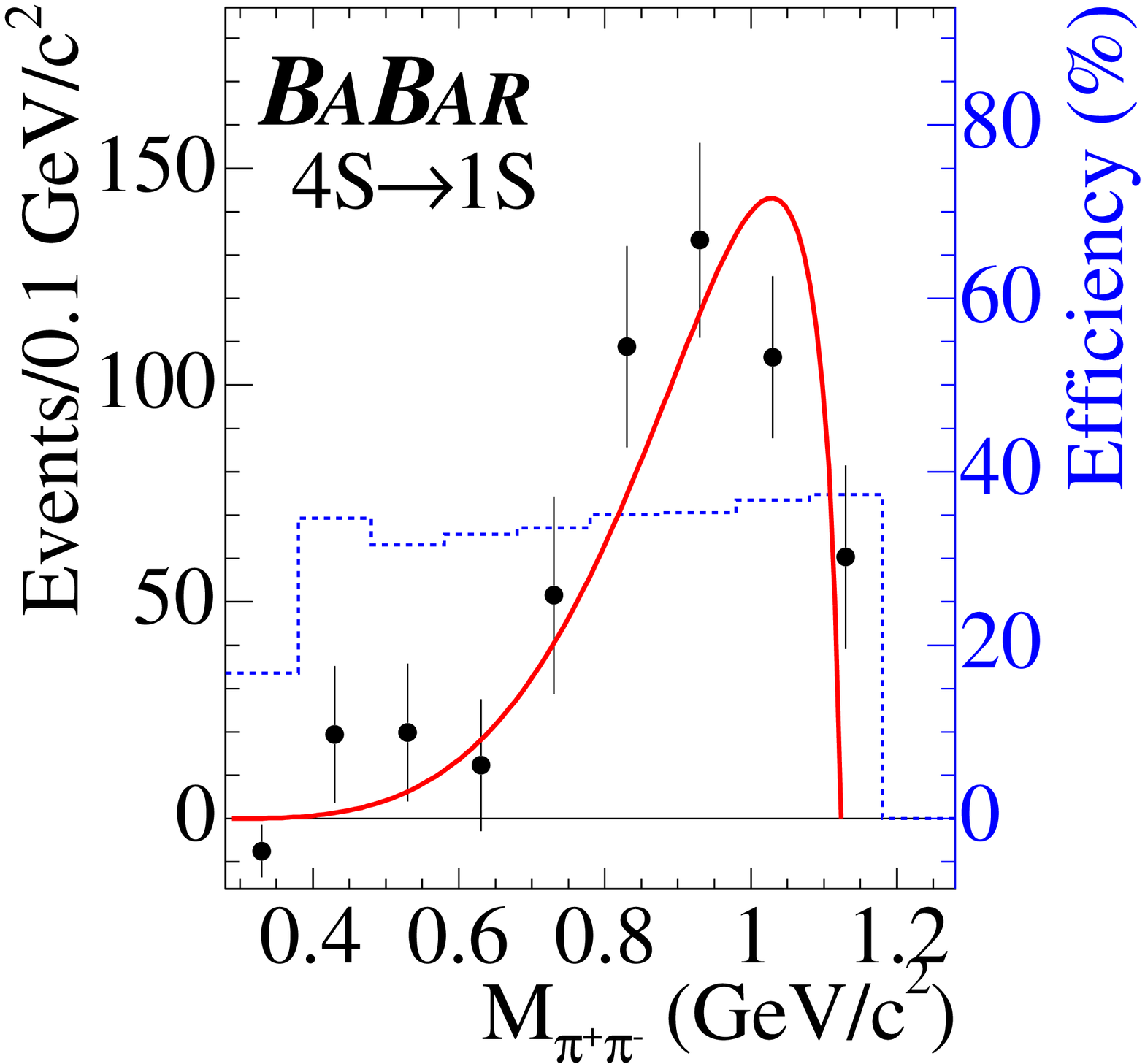,width=2.2in}
\epsfig{file=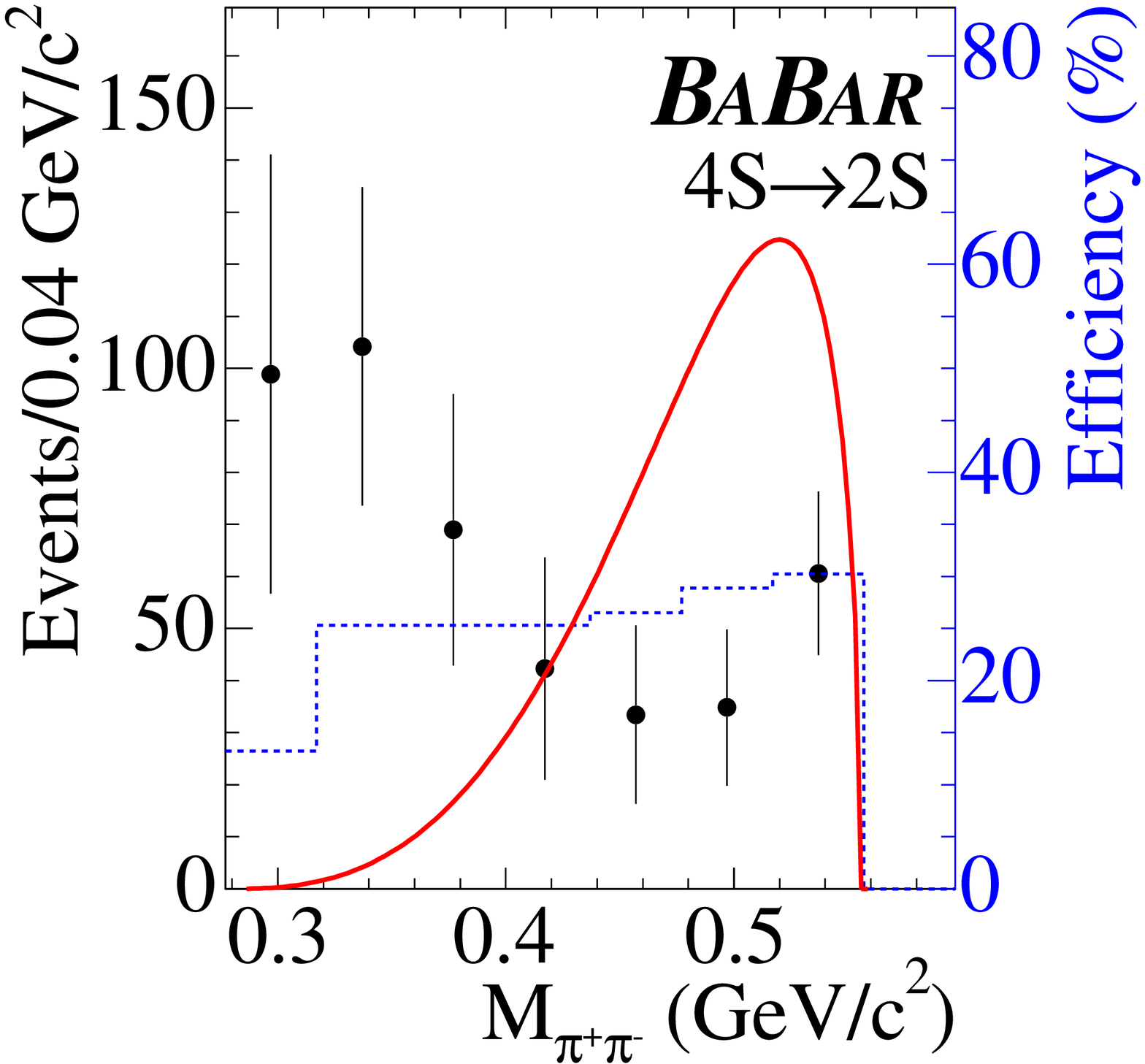,width=2.2in}
\caption{$\pipi$ invariant mass distributions in $\upsns\goesto\upsms\pipi$ from (top)CLEO 
$2S\goesto 1S\pipi$ and $3S \goesto 1S\pipi$, 
(middle) Belle $4S\goesto 1S\pipi$ and $3S \goesto 1S\pipi$~\cite{belle4s}
(bottom) BaBar $4S\goesto 1S\pipi$ and $4S \goesto 2S\pipi$.~\cite{babar4s}}
\label{allinvm}
\end{center}
\end{figure}

It has long been known that the invariant mass distribution of the $\pipi$ 
produced in $\upsiii\goesto\upsi\pipi$ is difficult to explain with a simple S-wave decay
model, but it is safe to say that the theoretical description of this decay remains incomplete.
CLEO has recently made measurements of the $\upsns\goesto\upsms\pipi$ transitions in the
most recent data sets,
and observed again the familiar double-humped structure in the invariant mass of the $\pipi$ 
produced in the $\upsiii\goesto\upsi\pipi$ transition, a preliminary plot of which is shown
in Figure~\ref{allinvm}.

BaBar and Belle have also made recent measurements of $\pipi$ invariant mass distributions 
in conjunction with the measurements discussed in the previous section (see Figure~\ref{allinvm}).   
Each of the three experiments
show $\upsiii\goesto\upsi$ and $\upsii\goesto\upsi$ distributions that agree with each other and
with the previously measured distributions.  Of particular interest, however, is the fact that the
$\Delta n = 2$ transition reported by BaBar, namely $\Upsilon(4S)\goesto\upsii\pipi$, also shows
a double-humped structure similar to the well-measured $\upsiii\goesto\upsi\pipi$ transition from CLEO.  
The fact that both these $\Delta n = 2$ transitions show similar $\pipi$ invariant mass structure
may be helpful in solving the long-known puzzle of the $\upsiii\goesto\upsi\pipi$ mass distribution. 

\section{Conclusions}

The past five years has been an active period in bottomonium spectroscopy, with
several new measurements contributing to the greater understanding of the
bottmonium system and heavy quark dynamics.  With the missing bottomonium singlet
states still to be found, and the question of a fuller explanation of hadronic
transitions within the system still to be answered, there is plenty of opportunity
for still more work to be done, should the opportunity arise. 


\section{ACKNOWLEDGMENTS}
The author thanks the International and Local Conference Committees for
their efforts at putting together a fabulous conference in a wonderful
setting.  The extensive preparations made for a most excellent visit, 
a smooth conference, and a thoroughly enjoyable time for all.  
The author also acknowledges the support of National Science Foundation
Grant No. PHY-0603831.


\begin{thebibliography}{99}


\bibitem{e288rept}
S. W. Herb \etal, \prl{39}{252}{1977}.
See also {\tt{http://history.fnal.gov/botqrk.html}} for a collection of very
intersting documents concerning the discovery.
\bibitem{qwg} Quarkonium Working Group Yellow Report, CERN 2005-005. See also {\tt{http://www.qwg.to.infn.it/}} for more information.
\bibitem{cleoee} J. L. Rosner \etal (CLEO), \prl{96}{092003}{2006}.
\bibitem{cleomumu} G. S. Adams \etal (CLEO), \prl{94}{012001}{2005}.
\bibitem{cleotautau} D. Besson \etal (CLEO), \prl{98}{052002}{2007}.
hep-ex/0607019 {\em Accepted for publication in PRL}.
\bibitem{pdg04}S. Eidelman \etal (PDG),\plb{592}{1}{2004} and partial
updates for the 2006 Edition available at {\tt {http://pdg.lbl.gov}}.
\bibitem{lgt}A. Gray \etal, \prd{72}{094507}{2005}.
\bibitem{gottfried}K. Gottfried, \prl{40}{598}{1978}. 
\bibitem{belle4s}K. Abe \etal (Belle), hep-ex/0512034, BELLE-CONF-510.  
\bibitem{babar4s}B. Aubert \etal (BaBar), \prl{96}{232001}{2006}.
\bibitem{babar4swidth}B. Aubert \etal (BaBar), \prd{95}{032005}{2005}.
\bibitem{pdg06}W.-M. Yao \etal {PDG}, {\em J. Phys. G}{\bf 33} (2006) 1.
\bibitem{newbelle}K. Abe \etal (Belle), hep-ex/0611026.
\end{thebibliography}
\end{document}